\documentstyle{cupconf}
\input{psfig}
\ifoldfss
\else
  \ifnfssone
    \newmathalphabet{\mathit}
      \addtoversion{normal}{\mathit}{cmr}{m}{it}
      \addtoversion{bold}{\mathit}{cmr}{bx}{it}
    \newmathalphabet{\mathcal}
      \addtoversion{normal}{\mathcal}{cmsy}{m}{n}
    \else
    \ifnfsstwo
    \fi
  \fi
\fi

\def\hexnumber#1{\ifcase#1 0\or1\or2\or3\or4\or5\or6\or7\or8\or9\or
 A\or B\or C\or D\or E\or F\fi }

\makeatletter
\ifx\CUP@mtlplain@loaded\undefined
\else
\fi
\makeatother
 \makeatletter
 \ifx\CUP@mtlplain@loaded\undefined
   \font\tenbmi=cmmib10 at 10pt
   \font\sevenbmi=cmmib10 at 7pt
   \font\fivebmi=cmmib10 at 5pt

   \newfam\bmifam
   \textfont\bmifam=\tenbmi
   \scriptfont\bmifam=\sevenbmi
   \scriptscriptfont\bmifam=\fivebmi
   
 \fi
 \makeatother
\ifnfsstwo

\fi
\ifnfssone

\fi
\ifoldfss

\fi

\mathchardef\varLambda="0103

\makeatletter
\ifx\CUP@mtlplain@loaded\undefined
\else
\fi
\makeatother
\makeatletter
\ifx\CUP@mtlplain@loaded\undefined
  \font\tenbms=cmbsy10
  \font\sevenbms=cmbsy10 at 7pt
  \font\fivebms=cmbsy10 at 5pt
  \newfam\bmsfam
  \textfont\bmsfam=\tenbms
  \scriptfont\bmsfam=\sevenbms
  \scriptscriptfont\bmsfam=\fivebms

  \edef\bsy@{\hexnumber\bmsfam}
  \mathchardef\bnabla="0\bsy@72
\fi
\makeatother
\def\eg{{e.g.\ }}

\def\etal{\mbox{\it et al.}}

\title[Cepheids as Distance Indicators]{Cepheids as Distance Indicators}

\author[N. R. Tanvir]%
{N\ls I\ls A\ls L\ns R.\ns T\ls A\ls N\ls V\ls I\ls R$^1$}

\affiliation{$^1$Institute of Astronomy, University of Cambridge,
Madingley Road, Cambridge, CB3 0HA, UK}

\begin{document}
\ifnfssone
\else
  \ifnfsstwo
  \else
    \ifoldfss
      \let\mathcal\cal
      \let\mathrm\rm
      \let\mathsf\sf
    \fi
  \fi
\fi

\maketitle

\def\PL{{PL}}
\def\PLC{{PLC}}
\def\H0{{$H_0$}}
\def\kms{km~s&^{-1}$}
\def\kmsmpc{km~s&^{-1}$~Mpc&^{-1}$}
\def\nb#1{{\bf (#1)}}

\def\sub#1{$_{_{#1}}$}
\def\Sub#1{_{_{#1}}}
\def\dm{$\mu\Sub{0}$}
\def\Dm{\mu\Sub{0}}
\def\mod#1{$\mu\Sub{#1}$}
\def\Mod#1{\mu\Sub{#1}}
\def\PW{{\PL\sub{W}}}

\def\imm#1{\raise0.4pt\hbox{$\langle$}$#1$\raise0.4pt\hbox{$\rangle$}}
\def\Imm#1{{\raise0.4pt\hbox{$\langle$}{#1}\raise0.4pt\hbox{$\rangle$}}}
\def\Square#1{
{\raise0.4pt\hbox{$[$}{#1}\raise0.4pt\hbox{$]$}}}
\def\log#1{${\rm log}(#1)$}
\def\Log#1{{\rm log}(#1)}

\def\lsim{{\small\mathrel{\hbox{\rlap{\hbox{\lower2pt\hbox{$\sim$}}}\raise2pt\hbox{$<$}}}}}
\def\gsim{{\small\mathrel{\hbox{\rlap{\hbox{\lower2pt\hbox{$\sim$}}}\raise2pt\hbox{$>$}}}}}
\def\ie{{\it i.e.}}
\def\eg{{\it e.g.}}
\def\kp{{\it key-project}}

\def\eqn{equation}

\begin{abstract}
We review the use of Cepheids as distance indicators
with particular emphasis on the methods which have been applied
to HST observations of Cepheids.
The calibration of the period-luminosity relations
is examined in detail and we identify possible
problems with the existing calibrations.
New $V$- and $I$-band period luminosity relations
are presented based on a sample of 53 Cepheids
in the LMC with photometry drawn from the 
literature.
These revised  \PL\ relations result in a
systematic decrease
of $\approx0.1$ magnitudes in distance moduli derived
using the standard method of extinction correction.
Hence estimates of $H\Sub{0}$ based on
such distances should be increased by $\sim5\%$.
Other aspects of Cepheid distance determination, specifically
incompleteness bias, metallicity dependence, the effects of crowding
and contamination of samples by non-Cepheids are
also discussed.
We conclude that current HST distance estimates to individual
galaxies are probably good to about 10\%, but that much of
this error is systematic.
Efforts to reduce the systematics, therefore, for example by improving the
photometric calibration, refining the distance to the LMC,
and reobserving the Cepheid galaxies in the
infrared with NICMOS, will give large returns.
\end{abstract}

\section{Introduction}

Cepheid variables are the most important primary distance indicators
and form the foundation of the extragalactic distance scale.
It was Henrietta Leavitt who in 
1912 first recognised their potential as standard candles
from her observations of variables in the Small Magellanic Cloud (SMC).
Subsequently \cite{Hubble24} used Cepheids to find distances to M31 and
M33, based on Shapley's (1918) period--luminosity (\PL) relation, thus 
proving the ``island--universe'' hypothesis of the nature of 
the spiral nebulae.
In 1952 Baade fundamentally revised the Cepheid distance
scale, differentiating for the first time between Cepheids of
populations I and II, with the result that distance estimates
to external galaxies were increased by
a factor $\sim2$ bringing them close to the modern values.
Recently the large investment of time in
Cepheid observations by the Hubble Space Telescope
(HST) has led to something of a renaissance of interest
in the field.
By the end of cycle 6 more than 20 new galaxies
will have been surveyed for Cepheids,
which it is hoped will lead to
the Hubble constant being established to better than 10\%.
It is now more important than ever to examine the
basis of Cepheid distance determination and to
rigorously evaluate the reliability of  Cepheid
distances.

Classical (population I) Cepheids are relatively massive stars, and hence
are short lived.
Consequently they are only found in late-type, spiral
and irregular, galaxies in regions of moderately recent star formation.
As distance indicators they have many desireable properties:
they are bright compared to most other stellar distance
indicators, reaching absolute visual magnitudes
of $M\Sub{V}\sim-6.5$; they are easy to identify from their
regular variability and characteristic light-curves; and the 
main distance
independent parameter, period of oscillation, can be obtained with
high precision.

Importantly Cepheids are also 
well understood theoretically. Briefly:
having exhausted its hydrogen core, a star
will evolve off the main-sequence and move 
rapidly across the Hertzsprung-Russell diagram
to the red-giant branch.
A sufficiently massive star will subsequently
begin to burn helium in its core and move back again
to higher temperature executing a ``blue loop'' (see \eg\ Chiosi 1990).
This second crossing of the HR diagram proceeds at a slower pace
and the star may spend a significant time in the
so-called ``instability strip''.
Possible causes of instability are many and complex (\eg\ Cox 1985), but
the major contributor in the case of Cepheids is thought to be the
He$^+$ ionization zone which, if it occurs at the appropriate
level in the star, will drive oscillations at the 
star's natural frequency
(or, in some circumstances, higher harmonics - see section \ref{contams}).
In fact, most of the variation in
luminosity is due to changes in temperature 
with only comparatively small changes in radius.
The natural frequency itself depends on the mean density
or, via the Stefan-Boltzmann law, on the mass, luminosity and
effective temperature of the star (Sandage 1958).
An implication of this is that the Cepheid \PL\ relation is
actually a projection of a more general period--luminosity--colour
(\PLC) relation, and the spread of the \PL\ is determined 
by the width
of the instability strip.

From the ground, 
reasonable samples of well-observed Cepheids
have only been obtained for galaxies at distances
$\lsim4$Mpc, which encompasses just the local group
and its few nearest neighbours.
With HST this distance range has been increased
to $>20$Mpc, representing an expansion of
more than 2 orders of magnitude in the available volume,
and bringing into play a number of rich groups
and clusters of galaxies.
Thus it is now becoming possible to calibrate
a host of secondary indicators using samples
of many galaxies which have direct Cepheid distances
and a much larger number which have distances by
association
(as witnessed by several other contributions to this proceedings).

In this review we concentrate mainly on the procedures
which have been used to obtain Cepheid distances with the
HST, and try to assess the remaining uncertainties
particularly the systematics.
These procedures are outlined in section 2
where we describe the means of correcting for reddening and their
consequences.
In section 3 we investigate the reliability of the 
period-luminosity relations in the $V$- and $I$-bands, and present
new calibrations based on a large sample of 53
Cepheids.
In section 4 we explore in some detail other potential problems
in using Cepheids, such as the effect of metallicity differences 
between the target galaxy and the calibrators;
the influence of
contamination of Cepheid samples  by overtone pulsators
and non-Cepheid variables; and statistical biases.
Section 5 examines some new developments which may
lead to improvements in Cepheid distance determination.
Finally, in section 6 we attempt to draw together
these threads to give a realistic estimate of the
full uncertainty on HST Cepheid distances.
There are a number of excellent 
reviews which also examine other issues: 
\cite{FW87}, \cite{MF91},
\cite{CL91} and Welch (in Jacoby \etal\ 1992).

\section{Cepheid studies with HST}

All Cepheid studies with HST to date have adopted essentially 
the same observing strategy.
$V$-band (WFPC2 F555W filter) images
are obtained at 12 or more suitably
spaced epochs spanning at least 8 weeks.
The spacing of the epochs can be optimised to 
search for variables in the range of interest,
typically $1<\Log{P}<1.8$ (see Freedman \etal\ 1994).
Photometry is then found for all stars on each frame
down to some magnitude limit.
Details of the standard photometric calibration are
given by \cite{Holtz95} and discussed further
by \cite{Hill96}, who also investigate the
so-called ``long--{\it vs}--short exposure''
correction, which is due to a small but poorly understood
non-linearity type problem with the WFPC2 chips.
These data are used to identify variables 
and measure their
periods.
Intensity mean $V$-band magnitudes, denoted \imm{V}
are calculated, usually with phase-weighting
as recommended by \cite{SH90}:

\begin{equation}
\label{sahamean}
\langle{m}\rangle=-2.5{\rm log_{10}}\sum_{i=1}^{n} 0.5(\phi_{i+1}-\phi_{i-1})10^{-0.4m_i}
\end{equation}
\smallskip

\noindent
where $\phi_i$ and $m_i$ are the phase and magnitude of the $i^{\rm th}$
epoch after folding on the best period.

Near infrared, $I$-band (WFPC2 F814W or WF/PC1 F785LP 
filters, which are close to Cousins $I$),
observations are also obtained, although at
fewer, typically 4, epochs. These data are combined with
the knowledge of the light curve shape and amplitude
from the  $V$-band to 
determine intensity mean $I$-band magnitudes, \imm{I}.
The transformation between $V$ and $I$ light curves
is discussed further in appendix A.

\subsection{Obtaining true distance moduli}

Usually the \imm{V} and \imm{I} magnitudes are combined
to estimate the reddening of the Cepheids themselves,
and hence the correction required to account for dust extinction.
This is done by fitting the $V$- and  $I$- \PL\ relations
separately to find apparent moduli,
\mod{AV} and \mod{AI},
and then calculating the true distance modulus for the sample
(\eg\ Freedman \etal\ 1994),\ie

\begin{equation}
\Dm=\Mod{AV}-R(\Mod{AV}-\Mod{AI})+({\rm correction~terms})
\label{method1}
\end{equation}
\smallskip

\noindent
Alternatively, we can estimate true distance moduli for each Cepheid
individually by \eqn\ \ref{method1}
and average the results (\eg\ Tanvir \etal\ 1995a; Saha \etal\ 1996),
We shall call this method 2 for future reference.
Although these two methods are mathematically identical,
method 2 can more easily
incorporate a weighting scheme
since it naturally handles
the fact that the residuals from the \PL\ relations 
in each band are correlated.
Note that here we define  $R=A\Sub{V}/(A\Sub{V}-A\Sub{I})$,
where $A$ is the extinction in the given band.
From the extinction curve of \cite{CCM89}, we find $R\approx2.45$,
is appropriate for Cepheids.
Other ``correction terms'', \eg\ for metallicity differences, 
are discussed below.

This general approach to 
the extinction problem has the benefit
that it corrects explicitly for the reddening to each
Cepheid based on the colour of the star itself.
Importantly,
the correction for extinction also takes out much of
the intrinsic PLC correlation between colour
and residual-magnitude of Cepheids
alluded to above.
Put simply, at a given period a Cepheid which appears
redder than the average may be so because it suffers from high
extinction or because it is towards the red edge of the
instability strip.  In both cases it will also appear fainter
than an average Cepheid would at that period and hence
its magnitude should be corrected brighter.
Similarly if a Cepheid is blue for whichever reason,
it is likely to be brighter than average.
The net result is that after applying the
reddening correction we are effectively dealing
with an intrinsically tighter relation, similar to the \PLC,
and hence obtain greater accuracy than would naively
be expected.
Rather ironically this means that the value of
the extinction, $A\Sub{V}$, is actually
{\em less} well determined than the value
of the true distance modulus, \dm.

It is interesting to note that this prescription is also 
essentially the same
as fitting an appropriate relation to 
the Wesenheit ``reddening-free magnitudes''
of the Cepheids, the $BV$ version of which has been investigated
in detail by \cite{Madore82} and \cite{Freedman88},
and which are discussed further here in section 3.
We should emphasize that there is no need for us to 
actually know the coefficients of the \PLC\ relation
with any precision in order to apply this method.
The fact that a \PLC\ relation exists and roughly
produces the same correlation between colour residual
and magnitude residual from the two \PL\ relations
as reddening, means that the 
reddening correction is doubly beneficial.

\subsection{The error budget}

Method 2 
provides an elegant way of
handling the errors since the spread in  the
sample of \dm estimates, which we shall call $\sigma_{\rm internal}$ ,
should be a fair estimate of all the
internal uncertainties due to random noise from photon statistics,
sampling of the light curves 
and the intrinsic dispersion (expected to be small as noted above).
This {\em automatically} accounts
for the correlation between the residuals from the $V$- and
$I$-band \PL\ relations which arise from the intrinsic PLC
and also the coupling of errors due to a common estimate of \log{P}.
So a fairly complete error estimate 
for a sample of $n$ Cepheids can be
written down simply by adding the various sources of systematic
error to this:

\begin{equation}
\sigma_{\rm total}^2=
\frac{\sigma_{\rm internal}^2}{n}+[R\sigma\Sub{I}]^2+[(R-1)\sigma\Sub{V}]^2+[\delta(\Mod{AV}-\Mod{AI})\sigma\Sub{R}]^2+\sigma\Sub{\rm PL}^2+\sigma\Sub{Z}^2+\sigma_{\rm systematic}^2
\label{budget}
\end{equation}
\smallskip

The two terms,
$\sigma\Sub{\rm PL}$, the uncertainty  associated with the
zero-point of the \PL\ relations themselves, and
$\sigma\Sub{Z}$, the uncertainty associated with any metallicity
correction, are discussed in more detail in sections
3 and 4 respectively.
The term $\sigma_{\rm systematic}$ is a catch-all and is intended
to account for any remaining systematic
uncertainties in the procedure for obtaining photometry such 
as acquiring
aperture corrections (\eg\ Tanvir \etal\ 1995b),
application of the ``long--{\it vs}--short exposure'' correction 
(Hill \etal\ 1996) etc.
Although at the moment these uncertainties are not well defined,
those which have been considered are expected to be small and
correlated between the bands and so should
not be a serious problem.
The observed interstellar extinction curve is fairly constant
for low density ISM and, in any case, the uncertainty in
the value of $R$ 
for the target galaxy won't be important
providing the estimate of colour excess 
{\em relative} to the LMC 
$\delta(\Mod{AV}-\Mod{AI})$
is not very high.
This is an advantage of observing in fields of reasonably low
extinction.

\begin{figure}[ht]
\centerline{\psfig{figure=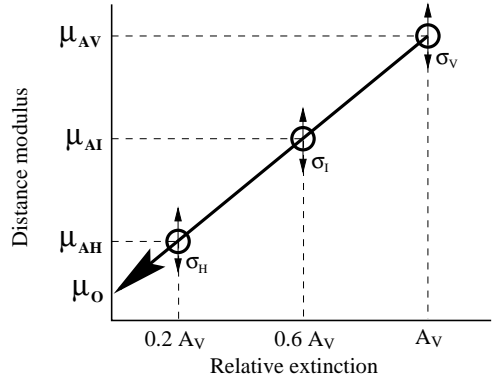,height=5cm}}
\caption{Schematic figure illustrating how the true distance
modulus may be obtained by combining the apparent distance
moduli in one or more bands.  The lever arm operates such
that for $VI$ observations photometric uncertainties, 
which propagate directly to uncertainties in the apparent
moduli \mod{AV} and \mod{AI}, are amplified in the error on
the true distance modulus.  
This produces both increased random scatter for individual
Cepheids, and importantly,  a large systematic uncertainty
due to the photometric zero-point errors.
By contrast, the addition of infrared
$H$-band magnitudes would result in a very well constrained value for
the true modulus.}
\label{leverarm}
\end{figure}

The main drawback of this reddening correction procedure,
as compared 
to having an independent estimate of the reddening, say, 
is that it is
more sensitive to {\em uncorrelated} photometric zero-point 
uncertainties, $\sigma\Sub{V}$ and particularly $\sigma\Sub{I}$.
This is illustrated schematically in figure \ref{leverarm}
which shows that the lever-arm for $VI$ observations
is large, although note that it is much smaller for infrared $H$-band
observations (see section 5).

Now the WFPC2 calibration uncertainty for $V$ and $I$ is
usually estimated to be 0.02--0.04 mag.
It is probable that there is also some zero-point uncertainty on
the ground-based measurements of the LMC Cepheids
in the region of 0.01--0.02 mag,
so we allow $\sigma\Sub{V}\approx\sigma\Sub{I}\approx0.04$ mag
for the {\em combined} photometric calibration uncertainty.
Thus these sources alone contribute a total error 
on the reddening corrected distance modulus of
$\sim0.11$ mag.
While this represents an uncertainty of only $\sim$6\%
in distance, it is important because
it is a systematic error affecting
all HST Cepheid distance determinations which use
this technique.

We shall return to \eqn\ \ref{budget} in
section \ref{conclusions} to bring together 
our estimates for the other terms.

\section{Calibration of the \PL\ Relations}

The standard \PL\ relations which have hitherto  been applied to HST Cepheids
are those of Madore \& Freedman (1991, hereafter MF91), based on
a set of 32 Cepheids in the Large Magellanic Cloud.
The advantages of calibrating in the LMC are that
the distance to the LMC can be obtained in different
ways, including by Cepheids which are calibrated
in the galaxy or using Baade-Wesselink
type methods. The SMC is less useful in this regard
because of its much greater depth along our line of sight.
A representative (but by no means exhaustive) selection
of recent LMC distance determinations is
given in table 1.
The question of whether there may be a small systematic discrepancy
between the RR Lyrae and Cepheid distance scales has been addressed 
most recently by \cite{vdB95}, \cite{Feast95} and 
\cite{Catalan96}.

\begin{table}
\begin{center}
\begin{tabular}{lll}
\quad Reference 
& \multicolumn{1}{c}{\dm} & \quad Method \\[4pt]
\cite{SOA84} & 18.2 & Main-sequence fitting \\
\cite{Walker92} & 18.22 & RR Lyraes calibrated by Milky Way RR Lyraes \\
\cite{Fernley94} & 18.43 & RR Lyraes using Baade-Wesselink method \\
\cite{Eastman89} & 18.45 & Expanding photosphere of supernova 1987A \\
\cite{Gieren94} &  18.47 & Cepheids using Baade-Wesselink (VSB method) \\
\cite{LS94} &      18.53 & Infrared Cepheid observations calibrated in Milky Way \\
\cite{Panagia96} & 18.54 & Ring around supernova 1987A\\
\cite{Feast95}     & 18.57 & Cepheids calibrated in Milky Way\\
\cite{HW90} & 18.66 & Mira variables calibrated against Miras in 47Tuc \\
\end{tabular}
\caption{Compilation of distance estimates to the Large Magellanic Cloud,
intended to be representative rather than exhaustive.
Although there is a fairly large range, the established techniques
mostly give distance moduli around 18.50 which corresponds to
50kpc.
}
\label{LMCtable}
\end{center}
\end{table}

MF91 adopted a true distance modulus to the LMC of $18.5\pm0.1$ and
a reddening of $E\Sub{B-V}=0.1$.  This distance modulus
still seems like a reasonable compromise between the different methods.
In fact, the adopted reddening is irrelevant to the derivation
of true distance moduli, if
they are calculated as described in section 2 (\eg\ Freedman \etal\ 1994).
The reddening to the LMC is, of course, still important
for several of the distance estimates in table \ref{LMCtable},
but we assume that the uncertainty on the reddening is included 
in the $\pm0.1$ error on the true modulus.

In addition to the uncertainty in the distance of the LMC, this
calibration will also propagate some uncertainty from the fit
to the given sample of Cepheids.  
In section 3.2 we present new \PL\ calibrations for which
the uncertainty on the fit is very small.
First we examine a potentially serious problem, 
identified by Simon \& Young (1996; hereafter SY96),
which leads us to ask the question whether we
have confidence in 
the LMC Cepheids as calibrators at all?

\subsection{Are the LMC Cepheids normal?}
\label{Simon}

Clearly the assumption underpinning the use of Cepheids as
distance indicators is that the Cepheids in a target
galaxy are similar to those in the calibrating galaxy,
in this case the LMC.
SY96 have pointed out 
possible differences from galaxy to galaxy in 
the distributions of long-period Cepheids in the colour-magnitude
diagram.
In particular they find the LMC Cepheids
to have a steep blue edge to the instability
strip when compared to the, apparently more normal, SMC Cepheids.
This, they argue,
can be explained if the LMC stars have 
a different mass-luminosity relation from the SMC.
Such variations in the M-L relation
could produce significant errors in distance modulus
of up to 0.25 mag when the LMC calibration is applied to another
galaxy with more ``SMC-like'' Cepheids.

\begin{figure}[ht]
\centerline{\psfig{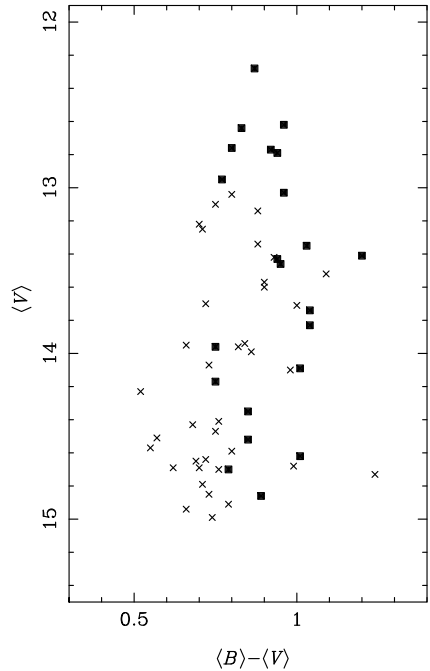}}
\caption{
Colour--magnitude diagram for LMC Cepheids with the
sample used by \cite{SY96}, namely those with $I$-band photometry
and periods in the range 10--50 days, 
displayed as solid squares, and Cepheids
not in their sample displayed as crosses.  We see that 
the unexpected vertical
slope of the blue-edge noted  by
Simon and Young is only an artifact of 
the unrepresentative
distribution of their sample compared to the whole population.
Presumably this is just an unfortunate consequence of small
number statistics, but in any event it is reassuring that the
larger sample shows no cause for concern with  the LMC
Cepheids.
These data were taken from the compilation of \cite{Madore85}.}
\label{cmd}
\end{figure}

But, the LMC Cepheid sample analyzed
by SY96 consists of just the 22 LMC Cepheids from
the \cite{Madore85} compilation which have
both $V$ and $I$ photometry and are in the period range 10 to 50 days.
The properties of this sample are investigated
in figure \ref{cmd} which shows the \imm{V} {\it vs} 
\imm{B}$-$\imm{V} colour-magnitude
diagram for all the LMC Cepheids 
from \cite{Madore85} in the period
range 5 to 50 days, highlighting the Cepheids
used by SY96.
We see immediately that these Cepheids are
not a fair sample of the full population.
In particular,  bluer Cepheids are
under-represented in the magnitude range \imm{V}$>13$ mag.
Presumably this is just a consequence of having such a small
numbers of variables. A simple-minded analysis suggests that
this offset of the 22 Cepheid sample from the mean of
the larger set is only
a $1\sigma$ event, although this increases to more than 
$2\sigma$ if, {\it a posteriori}, we consider only the
11 Cepheids with $\Log{P}<1.45$ for which the discrepancy is
most pronounced.
In any case we conclude that the effect
noted by SY96 is due to an unfortunate anomaly
of their sample.

This is reassuring for the use of the Cepheid \PL\ relations
in general since the evidence for a varying M-L goes away.
However, these same 22 Cepheids form a
large proportion of the sample   used by MF91 in 
calibrating their \PL\ relations, and are also important,
but to a lesser extent,
in the new calibrations presented in this paper in section 3.2.
It is therefore relevant to ask how bad a calibration
would have been constructed using these 22 Cepheids alone?

In this context the sampling problem manifests itself as an
deficiency of Cepheids in the range  $1<\Log{P}<1.45$ which
lie above the mean \PL\ relation. The result is that
both the estimated slopes and zero-points
of the \PL\sub{V} and \PL\sub{I} relations will be significantly
in error.
This unusual slope can be seen, incidentally, in most direct
comparisons of LMC Cepheids with those in other galaxies
(\eg\ see figure 10 of Freedman \etal\ 1994, or
figure 7 of Kelson \etal\ 1996).
For distance determination, the consequences of this 
depend in general on the period distribution of the Cepheids
in the particular target galaxy.
However, even for a ``bad'' case, where the target
Cepheids are concentrated around $\Log{P}=1.2$,
we find that the estimate of true distance modulus
by \eqn\ \ref{method1}
is only out by 0.01--0.03 mag for either forward or reverse
regression fits to the data.
This is a remarkably small error and
again comes down to the small intrinsic dispersion of the
reddening corrected magnitudes.
Inevitably, the estimate of reddening itself is worse
affected and can be out by as much as 0.1 mag, but this just illustrates
the point made earlier,
that the estimate of true distance modulus should be taken
more seriously than the estimate of the reddening.
In fact for this reason, and also because (a) the nominal
$E\Sub{B-V}$ reddening to the LMC Cepheids is itself uncertain,
and (b) the reddening is also affected by photometric errors,
we should not be afraid to apply a ``bluening'' correction
if that is what is found.  Indeed if we forbid negative reddening,
at least if we are to work consistently
within the framework described in section 2,
then we run the risk of creating a bias on the resultant distance
estimates.

The result of calibrating with the 22 Cepheid
sample would have been considerably worse had we  departed from
the ``standard'' method.
If, for example, we have an external estimate of the
reddening not derived from the $VI$ photometry,
then the 
error in distance modulus would have been
$\sim0.14$ mag.
If we had adopted a \PL\sub{V} relation based on the larger
sample of Cepheids with $V$-band photometry, and used this
in conjunction with the 22 Cepheid \PL\sub{I} relation to
correct for reddening, then an error in distance modulus
of 0.3 mags would have resulted.

\subsection{New calibrations of the \PL\ relations}

\begin{figure}[pth]
\centerline{\psfig{figure=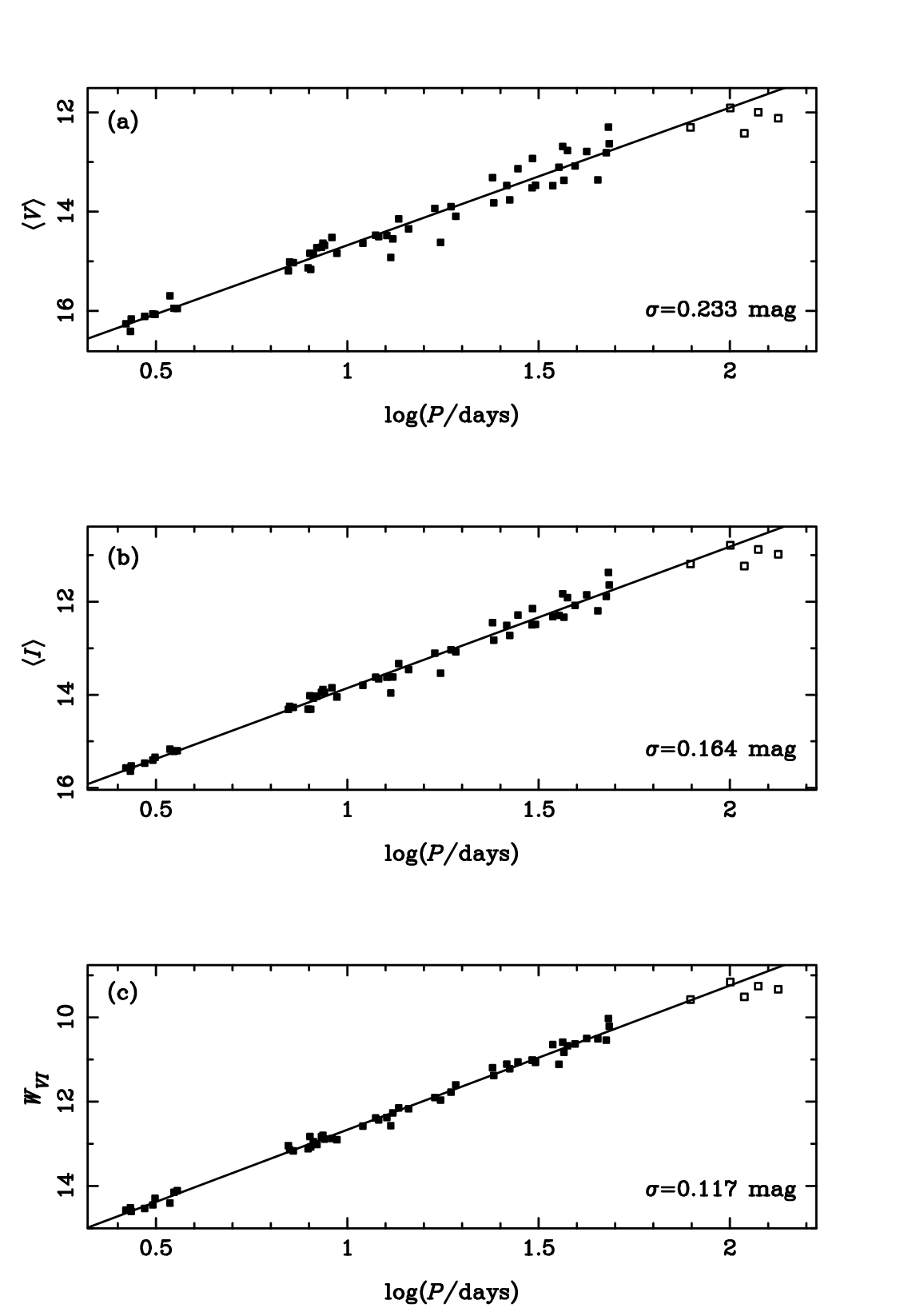,height=15cm}}
\caption{Period-luminosity relations for a sample of 53 LMC
Cepheids (solid squares) and 5 long period Cepheids
(open squares)
with $V$- and $I$-band photometry from the literature.
Data were taken from \cite{Madore75}, \cite{Eggen77}, \cite{MW79}, 
\cite{Martin80}, \cite{Dean81}, \cite{vG83}, \cite{Harris83}, \cite{FGM85},
\cite{Walker87}, \cite{Welch91}, \cite{CCSBF86}, \cite{Gieren93}
and \cite{SW95}.
The magnitudes for the 7 Cepheids from \cite{SW95} were used as given.
Many other Cepheids turned out to have been
studied by more than one group, so
those data were combined and obviously
discrepant data sets and data points removed by hand.
The three panels are for (a) 
intensity mean 
\imm{V} magnitudes which were calculated according to \eqn\ \ref{sahamean},
(b) intensity mean \imm{I} magnitudes which were calculated using
$\Imm{I}=\Imm{I}^{\prime}+0.6\left[{\Imm{V}-\Imm{V}^{\prime}}\right]$,
where the primes indicate that the intensity means (not phase weighted)
were taken using
just those epochs which had both $V$- and $I$-band photometry
(see appendix A for justification of the coefficient 0.6),
and (c) $W\Sub{VI}$, the reddening
free Wesenheit function (\eg\ Madore 1982) which is defined here
as $W\Sub{VI}=\Imm{V}-2.45\Square{\Imm{V}-\Imm{I}}$.
The solid lines are least-squares fits to the Cepheids
with $0.4<\Log{P}<1.8$.  Note the small dispersion
around the mean relation, particularly for $W\Sub{VI}$,
and the absence of deviation from linearity.
The apparent tendency for points to scatter preferentially
below the mean line
in the period range $1<\Log{P}<1.4$ is explained in section \ref{Simon}
and has no significant influence on the calibrations.
}
\label{3PLs}
\end{figure}

Several hundred Cepheids have already been observed in
distant galaxies with HST and it is clearly desirable
to increase the size of the LMC $VI$ sample
to minimize any systematic error on the calibration.
With this in mind
we have re-examined the
published $V$ and $I$ photoelectric photometry for LMC Cepheids.
By combining all the available data sets (post--1975) we obtain a sample
of 53 Cepheids.  The period-luminosity relations for
these Cepheids are shown in figure \ref{3PLs}.
We also plot the Wesenheit magnitude
(see Madore 1982)
which is constructed from the
$VI$ photometry,
$W\Sub{VI}=\Imm{V}-R\Square{\Imm{V}-\Imm{I}}$,
and is explicitly reddening independent.
Notice that the Wesenheit magnitudes
produce a very tight \PL\ relation with a
dispersion of $<0.12$ mag.
As noted above, this tightening is expected as 
a result of the fact that the
intrinsic PLC relation works in the same
sense as reddening.
Nonetheless, such a low dispersion seems remarkable given
the inhomogeneous collection of data sets used in creating
the sample, and recalling also that the dispersion must
include contributions from random
photometric errors and the depth of the LMC in addition to the
intrinsic scatter around the relation.
On this point, we might expect the
scatter to improve
still further if we
remove the effect of the tilt
of the LMC in the plane of the sky, as determined by \cite{CL91}.
In fact, doing this
we find a small reduction in 
the dispersion around the $V$- and $I$- \PL\ relations,
to 0.226 and 0.159 mag respectively, but essentially no change in
the dispersion for the $W\Sub{VI}$ relation.  This probably
indicates that photometric errors are becoming dominant.

If we follow MF91 and adopt a distance modulus
to the LMC of $\Dm=18.5$ and reddening $E\Sub{B-V}$=0.1
(and therefore $A\Sub{V}=0.33$ and $A\Sub{I}=0.195$), 
we obtain the following calibrations:

\medskip

\begin{equation}
\fbox{\shortstack{
$M\Sub{V}=-2.774 (\pm0.083)\Square{\Log{P}-1.4} - 5.262(\pm0.040)$\qquad ;\quad$\sigma_{rms}=0.233$ \\
$M\Sub{I}\ =-3.039 (\pm0.059)\Square{\Log{P}-1.4} - 6.054(\pm0.028)$\qquad  ;\quad$\sigma_{rms}=0.164$ \\
$M\Sub{W}=-3.423 (\pm0.042) \Square{\Log{P}-1.4} - 7.202 (\pm0.020)$\qquad ;\quad$\sigma_{rms}=0.117$ 
}}
\label{Box}
\end{equation}

\smallskip
\bigskip
\noindent
which are fits to all the variables with $\Log{P}<1.8$,
avoiding potential problems with the longest period Cepheids
(MF91), and are
referenced to a pivot $\Log{P}$ of 1.4, typical of the
HST Cepheid samples.
Note that the uncertainty on the zero-point of the fitted \PW\ relation
is only 1\% in distance.
	
\begin{figure}[th]
\centerline{\psfig{figure=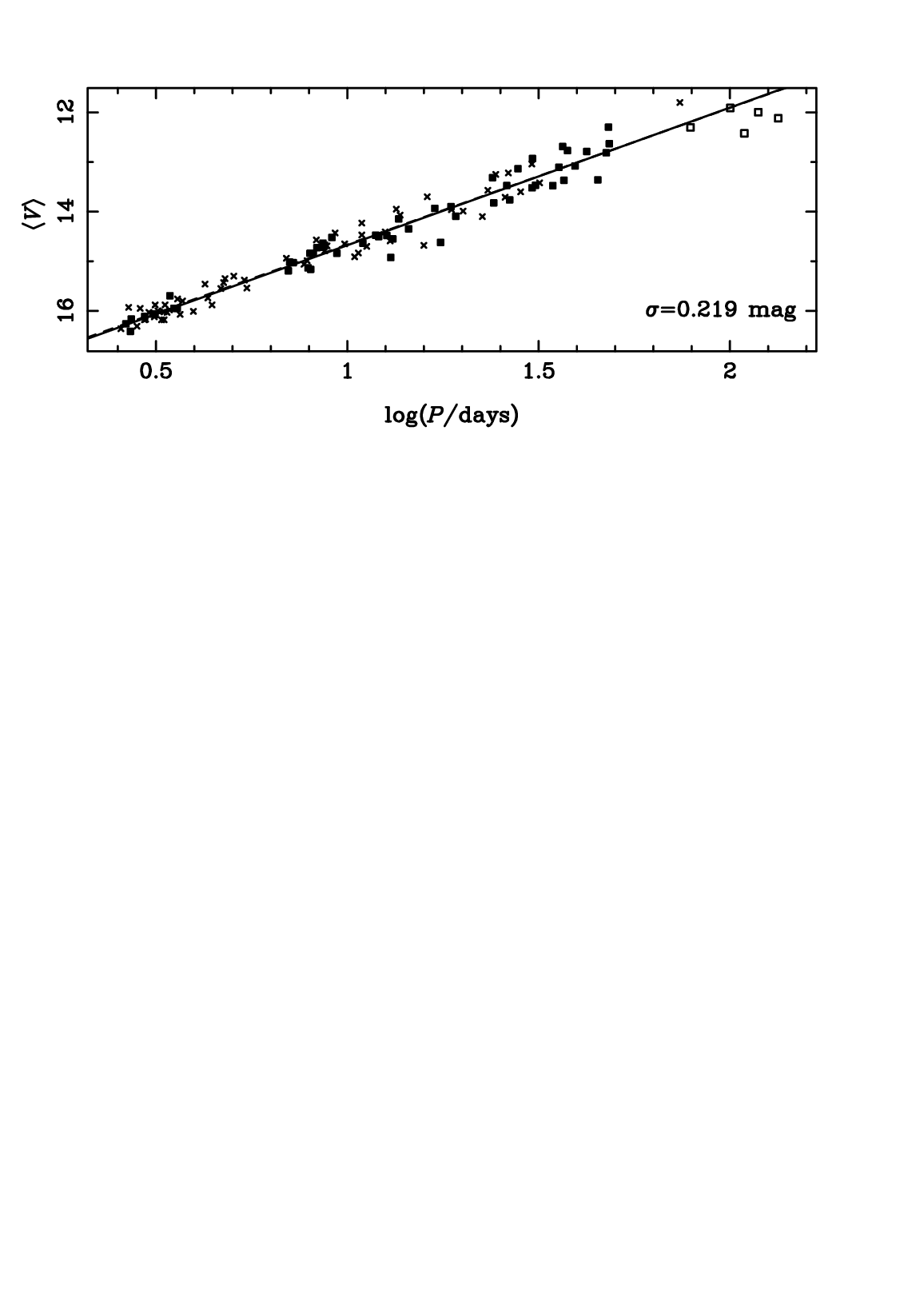,height=5cm,bbllx=0pt,bblly=527pt,bburx=543pt,bbury=790pt}}
\caption{The period-luminosity plot for the sample of 53 LMC
Cepheids (plus 5 with $\Log{P}>1.8$) which have both $V$- and 
$I$-band photometry are again shown as solid squares 
and solid line, as in figure 
\ref{3PLs}a.
This is compared to a sample of 58 
Cepheids (plus 1 with $\Log{P}>1.8$) with $V$ but no $I$ 
photometry (crosses) which have been compiled from:
\cite{Martin81}, \cite{vG89}, \cite{Mateo90}, \cite{Bertelli93}
and \cite{Welch93},
in addition to the sources referenced in the caption to 
figure \ref{3PLs}.
Variables identified as overtones and the highly 
reddened Cepheid HV2549 have not been included.
The least-squares fit for the entire sample of 111 variables
(dashed line)
is barely distinguishable, particularly in the period range,
$\Log{P}>1$,
of interest for distant studies.
}
\label{1PL}
\end{figure}

We may ask whether even these 53 Cepheids form a fair sample.  To test this
we plot in figure \ref{1PL} an additional  
sample of 58 Cepheids with $V$-band
but no $I$-band photometry, also drawn from the literature.
From the figure we see that while 
there is still some deficiency in the number
of brighter Cepheids with 
$I$-band photometry
in the range $1<\Log{P}<1.4$, nonetheless,
the fit to the combined sample of 111 Cepheids
is extremely close to the fit
for the 53 Cepheid subset:

\begin{equation}
\hbox{$M\Sub{V}=-2.756 (\pm0.054)\Square{\Log{P}-1.4} - 5.269(\pm0.031)$\quad ;\quad$\sigma_{rms}=0.219$}
\end{equation}
\smallskip

Restricting to a subset of 74 
longer period variables ($1.0<\Log{P}<1.8$)
leads to an almost identical relation:

\begin{equation}
\hbox{$M\Sub{V}=-2.780 (\pm0.084)\Square{\Log{P}-1.4} - 5.263(\pm0.035)$\quad ;\quad$\sigma_{rms}=0.245$}
\end{equation}
\smallskip

\noindent
which shows that the LMC \PL$\Sub{V}$ relation is {\em linear},
to this level of precision, and that the version given
in \eqn\ \ref{Box} is {\em representative}.
Thus there is no compelling reason to restrict the period
range of the calibrating sample to be the same as the
period range in any particular target sample.

For comparison, recent
studies of galactic Cepheids using both Baade-Wesselink methods
and in clusters with main-sequence distances
have tended to give somewhat steeper $V$-band \PL\ slopes, usually
in the
range 2.9--3.0 (see Gieren \etal\ 1993).
The theoretical metallicity dependence found by
\cite{CWC93} (and summarized here in \eqn\ \ref{ZVPL})
predicts that there should be an
increase in slope, but by only  $\sim0.06$,
however this is quite consistent within the errors.

\subsection{Implications of the new calibrations}

Our \PL\sub{V} relation agrees well with MF91, who
found $M\Sub{V}=-2.76\Square{\Log{P}-1.4} - 5.27$,
indicating that the lack of brighter Cepheids in the
range $1<\Log{P}<1.45$ does not have much influence on
their fit either.
However there is disagreement in the $I$-band, where
MF91 gave $M\Sub{I}=-3.06\Square{\Log{P}-1.4}-6.09$.
The chief reason for the discrepancy can be traced back to
the $V-I$ colours listed by \cite{Madore85}
for the \cite{MWF79} Cepheids, which are actually
straight magnitude means, \ie\ $\overline{V-I}$.
The \imm{I} magnitudes going into the MF91 calibration
appear to have been
estimated by subtracting these colours from the \imm{V}
magnitudes, and this produces small but 
systematic errors
of 0.02--0.09 mag for Cepheids of normal amplitudes.
Thus the difference between the two calibrations stems
largely from the re-analysis of the photometry
for the variables we have in common, rather than the
increased sample used here.

The net result is that distance moduli calculated with MF91
relations, using the $V$ and $I$ results to give the reddening
via \eqn\ \ref{method1},
should be {\em reduced} by $\approx0.1$ mag.
Note how the small revision in the $I$-band zero-point
is magnified in the reddening correction procedure.
This reduction in distance 
translates to a $\sim5$\% {\em increase} in inferred values
of the Hubble constant.
Although small, this revision is significant given
the hoped for level of precision in determining \H0.
We should point out that for some of the Sandage \etal\
SNIa host galaxies, for which it was assumed that the
reddening to the Cepheids is the same as that to the supernova
(eg. Saha \etal\ 1994), the effect on the \H0\
determinations will be less.

In passing we also note that the ground-based Cepheid distances
to more nearby galaxies
will be less seriously affected since none rely on
only $VI$ photometry to give reddening.

\section{Some other concerns in applying Cepheids}

Apart from the calibration questions dealt with in 
section 3, there are a number of other issues
we should address regarding the application
of Cepheids in other galaxies.
To preempt the gory details, a common conclusion of
the first three topics is that imposing a reasonably
conservative lower \log{P} limit on a Cepheid sample
will help remove potential systematic problems
which can appear at faint magnitudes, but will not usually
limit the accuracy of the distance estimate.

\subsection{Statistical biases}

Much has been written about the effects of statistical
biases on Tully-Fisher and $D_n-\sigma$ distance estimates
(\eg\ Hendry \& Simmons 1994; Triay \etal\ 1994)
but less attention has been paid to the equivalent problems
for Cepheids.
\cite{Sandage88} demonstrated that Cepheid samples
which are truncated by a detection limit do indeed
show flatter \PL\ slopes than
complete samples and hence produce biased distances.
The affect of this, and other sample selection
effects, has been analyzed by \cite{HK96} for 
Sandage \etal's HST $V$-band
Cepheid observations of NGC5253 and IC4182.
They identify bias at a low level, but
conclude that it is not a significant source of error
in determining \H0.
\cite{Feast95} pointed out that 
applying reverse rather than direct regression
in some circumstances
helps avoid problems of this kind.
This argument was also made
by \cite{Kelson96} who adopted
reverse fitting in determining the HST
distance to M101.

There are some issues of principle here, though,
which should give us pause for thought.
The LMC Cepheid calibrating sample 
as it stands is limited
in {\em period}, not magnitude, and indeed several of the
studies of LMC Cepheids from which the photometry is
derived have selected samples in 
particular period ranges.
Furthermore, the period distribution of Cepheids in
a given galaxy will depend at some level on its
star formation history, and
the efficiency with which variables will be found
in the target galaxy will be a function of period
due to the particular spacing of the observations.
Additional sources of scatter, such as differential
reddening and photometric errors, will work  to
broaden the distribution in magnitude
and will therefore also affect the reverse fit
given the various forms of period selection
we have identified.
If we assume that the only important period selection
effect is the upper cut off, which of course
applies to both the LMC and the target galaxy,
then it may be possible to proceed by imposing a magnitude
selection function on the LMC sample which is at least
close to that for the target galaxy.
The alternative, of imposing a bright magnitude 
limit on both samples seems rather wasteful.
In light of these concerns, we feel a better
policy, if one is needed, would be to continue with forward
regression but to impose a short period cut-off on
the target galaxy sample to remove any range in
\log{P} which appears to be badly incomplete in magnitude.

\begin{figure}[ht]
\centerline{\psfig{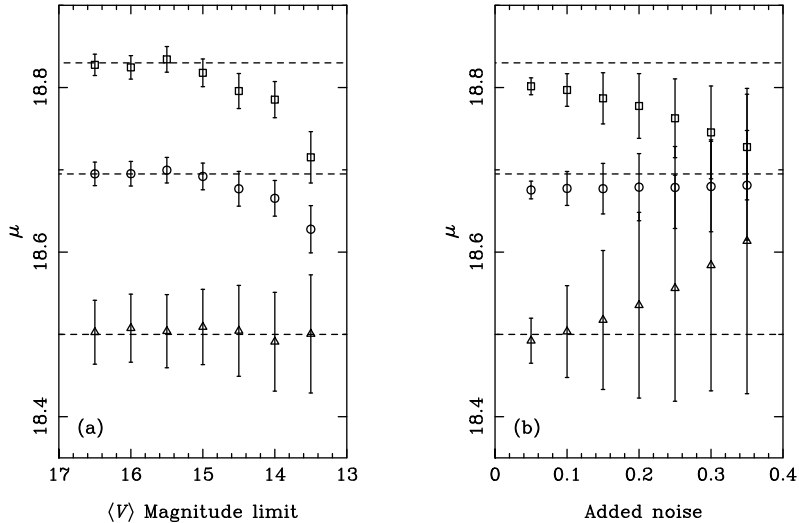}}
\caption{This figure illustrates the effects of
incompleteness bias in the presence of noise.
The \PL\ relations given in \eqn\ \ref{Box} are applied
to artificially degraded LMC samples and the results
for \mod{AV} (boxes), \mod{AI} (circles) and \dm 
(triangles) can be compared to the
nominal values (dashed lines).  In panel (a) the
photometry for the entire sample, $V$ and $I$, has
been degraded by the addition of 0.1 mag of random
noise, to simulate the higher measurement errors
for the distant Cepheid samples, 
and the bias is created by introduction
of a sharp artificial $V$-band magnitude limit.
The error bars show the distribution over different
realizations.
This demonstrates that while \PL\sub{V} and \PL\sub{I} alone
can give biased results from severely incomplete samples,
the reddening corrected modulus, although noisy, is
not biased.
In panel (b) we instead fix the magnitude limit at
$\Imm{V}=14.5$, which corresponds to $\Log{P}\approx1$
typical of the distant Cepheid surveys,
and plot the recovered distance moduli 
as a function of the
added noise.  The purpose of this is to show
that random, uncorrelated (between $V$ and $I$)
photometric noise increases the bias in
determining \mod{AV}, but actually
produces a bias
on the
reddening corrected modulus
which is in the opposite sense!
The effect of correlated noise depends on the
nature of the coupling, but will usually
bring the bias on \dm down again.
The conclusion is that statistical biases 
can produce complex effects, but are 
likely to be small for realistic levels of
photometric noise.
}
\label{bias_sims}
\end{figure}

But does any of this matter in practice?
Distance indicators with large dispersion are
more prone to selection biases,
thus \PL\sub{V} used alone, for example, will produce
distance estimates which are systematically too
short in the presence of a detection limit.
However, we have seen that the reddening
corrected $W\Sub{VI}$ magnitudes form a linear \PW\  
relation with very small {\em intrinsic},
scatter,
so as an indicator \PW\ 
should be much less biased than
\PL\sub{V}.
Random photometric noise will of course broaden
the relation, but because sample selection
is done basically on the \imm{V} mags,
rather than on \imm{I} mags, it turns out that
net bias for \dm\ estimates
actually starts to go in the opposite
sense, \ie\ upward.
On the other hand random but {\em correlated}
source of error
for each Cepheid, which will be present since the $I$ 
light curve
uses information from the $V$ light curve and crowding
errors affect both bands,
can move the bias down again as with \mod{AV}.
We demonstrate these points in figure \ref{bias_sims}
where we assume the relations given
in \eqn\ \ref{Box}, and apply them to the original
LMC sample degraded by noise and a sharp artificial
magnitude limit.
Although not entirely realistic, these experiments
serve to illustrate that selection effects can produce
complicated biases, but
overall, for typical levels of noise,
residual bias should be small.
This explains why \cite{Kelson96} did not
find the difference between forward and
reverse fitting in M101 that they were expecting.

In conclusion, then, normal slide fitting of the forward
\PL\ relations should not produce any strong bias due to
incompleteness, although if the random (uncorrelated)
photometric noise near
the magnitude limit is high then distances
can be biased too far.
More thorough appraisal of selection biases would require
realistic simulations of artificial Cepheids being
added to the images and
put through the same reduction procedure as the real data.
On a different but related issue,
Malmquist bias, used here in the sense of the bias resulting from the
non-constant radial 
distribution of the target galaxies,  will depend in general 
on the scatter, or internal
error if you like, of the Cepheid distances.
Since these errors are usually small, compared to indicators
like the Tully-Fisher  and $D_n-{\sigma}$ relations,
Malmquist bias should also be minimal.

\subsection{Contamination by overtone pulsators and non-Cepheids}
\label{contams}

Could there be contamination
of the distant samples
by Cepheids pulsating in higher harmonic modes or by
variables which are not classical Cepheids at all?
Such contamination could produce  systematic
errors in distance determination.
In fact, the only plausible non-Cepheid interlopers which might
appear in the same colour, magnitude and period range
with anything like Cepheid light curves,
are the  W Virginis stars.
These variables, also known as population II Cepheids,
follow a \PL\ relation about 1.5 mags fainter than
the classical Cepheid relation.
However, the statistics of variables 
discovered in ground-based surveys of the Magellanic Clouds, M31 and 
NGC300 , compiled by
\cite{MF85}, show that  
whilst more than 70\% were classical  Cepheids,
only $\sim$2\% were W Virginis stars, making serious contamination
unlikely.
A further test was performed by \cite{SC93}
who compared the light
curve shapes of 5 Cepheids in IC4182 
(from Saha \etal\ 1994) 
with galactic Cepheids
using Fourier decomposition.
They concluded that the samples are the same within the
errors and, in particular, there is no indication that any of
their IC4182 Cepheids are in fact W Virginis stars.

Cepheids pulsating in first harmonic mode on the
other hand should follow
a \PL\ relation which is displaced to lower period
and hence apparently to brighter magnitude
than the \PL\ relation for fundamental mode Cepheids.
\cite{BV94} has suggested that a 
majority of classical Cepheids with periods less than $\Log{P}\lsim0.9$
are 1H pulsators.
Overtone pulsators may be distinguished on
the basis of their light curves, although this would
be difficult in any quantitative sense for the relatively poorly
sampled, noisy light curves of Cepheids in distant galaxies.
Nonetheless, strongly saw-tooth light curves 
provide some reassurance that the variables are classical
Cepheids pulsating in fundamental mode.

In fact, the results from the microlensing surveys of
the LMC appear to show that while overtone pulsators
do exist with periods as long as $\Log{P}=0.8$
they form a relatively small proportion of all Cepheids even at
$\Log{P}=0.5$ (see figure 3 of \cite{MACHO95}
for results from the first year of MACHO data).
Moreover, given that most HST samples are restricted to
$\Log{P}>0.8$ anyway, simply because of the magnitude limit
of the observations, contamination by overtone pulsators
is unlikely to be a problem.

Finally we mention a category of non-Cepheid 
interloper which is
frequently overlooked, namely the non-variable stars!
At sufficiently faint magnitudes, given
the relatively small number of epochs used,
random photometric errors will cause some stars
to appear as ``variable'' as true variables.
These usually will not fold to produce 
Cepheid light curves and hence will be rejected,
but it is possible that some may slip through
if the search for variables is pushed to faint
enough magnitudes and low enough amplitudes.
The scale of this effect for any particular field 
can be best judged by
simulations.
Here we simply make the point again that 
a conservative cut in \log{P}
will exclude the fainter ``variables'' 
for which there is less assurance that they are
genuine Cepheids.

\subsection{The effects of crowding}
\label{crowding}

The photometry for many HST target galaxies is  made difficult
by stellar crowding and rapid variations in background intensity.
Profile fitting photometry software has been
shown to work quite well despite the poor
sampling of the WFPC2 CCDs.
Such algorithms attempt to use knowledge of the point-spread-function
to fit for the magnitudes of many stars simultaneously,
thus effectively correcting for the influence of
light from the wings of neighboring stars.

However, in severe cases problems are bound to arise.
The \kp\ group have found some 
discrepancies at a fairly low level between
the results obtained with the DAOPHOT/ALLFRAME software
and those obtained with the 
DoPHOT software in the rather crowded M100 fields (Hill \etal\ 1996).
\cite{Sahaetal96} have investigated this problem in NGC4536 by assigning
a ``quality index'' to each variable based on an inspection of the light
curves.  They find that variables with a low quality index
are more likely to be recovered with brighter magnitudes at
a given period, which may well indicate the effects of crowding.
Indeed crowding errors clearly become dominant
for $\Log{P}<1.2$ in their data,
especially in the $I$-band.

Interestingly, if the photometry for a Cepheid is contaminated by the
light of an overlapping blue main-sequence star
then the reddening correction (\eqn\ \ref{method1}) again works in the 
sense to minimize its influence.
Contamination by young blue stars is not improbable given
that Cepheids tend to be found in regions of moderately
recent star formation.
Conversely, of course, contamination
by red stars can produce larger errors in the
reddening corrected distance estimates.
The tip of the red giant branch for an old population has
$M\Sub{I}\approx-4$ so Cepheids with $\Log{P}\lsim0.8$ 
will be more subject to contamination by such stars.

Experiments with adding simulated stars to the
data frames should give a good idea of the scale of
the problem in a particular galaxy,
and may allow some form of empirical correction to the magnitudes to be
estimated.
A signature that crowding errors are becoming dominant would be
if the distance moduli of the individual Cepheids (using method 2)
show a trend with period in the sense of short period
Cepheids appearing systematically closer.
Possible ways to minimize the effects of crowding are to
(a) reject any stars which appear at all broader than the psf, (b)
cut the sample in period so as to only include the 
longer period Cepheids, or
(c) cut according to amplitude since contamination by a non-variable
star will
always result in a reduction in the amplitude.
Given the small intrinsic dispersion of the Wesenheit
magnitudes, cutting the sample size will not be 
detrimental for the distance estimate,
but including faint Cepheids with biased {\em photometry},
due to crowding could be.

\subsection{Differences due to metallicity}
\label{metals}

In principle metallicity may affect the Cepheid \PL\ relations
due to changes in stellar evolution, pulsation or
atmospheres.
Theoretical studies (\eg\ Stothers 1988; Chiosi, Wood \&
Capitanio 1993, hereafter CWC93)
suggest that metallicity effects are very important
in the $B$-band, but less so in $V$ and $I$.
For example, using the results given by CWC93 we compute, 
for a change in metallicity $\delta Z$, 
the measured
apparent $V$-band distance modulus will change by:

\begin{equation}
\delta\Mod{AV}=\Square{7.73\Log{P}-0.49}\delta Z
\label{ZVPL}
\end{equation}

\smallskip

\noindent
where we are additionally assuming that
(a) the helium abundance goes as $\delta Y=3.5\delta Z$ (Peimbert 1986),
(b) there are no changes in convective overshoot properties
with metallicity (the CWC93 models fit the available Milky-Way, LMC
and SMC
data best if overshoot is zero or small), 
and (c) the Cepheids more or less uniformly
fill the instability strip, which may be the
weakest assumption particularly given the rather uncertain
prescription for defining the red-edge of the strip.
Anyway, this amounts to $0.08$ mag for a Cepheid sample
centred around $\Log{P}=1.4$ and $\delta Z=0.008$ which is appropriate
for the difference between LMC and galactic metallicity for example.
In other words, if we measure the distance to a target galaxy
of galactic metallicity using the LMC Cepheids as
calibrators, then we should apply a correction to the distance
modulus of $-0.08$ mag.
However the method of extinction correction outlined in
section 2 actually produces a cancelation of most of this dependency:

\begin{equation}
\delta\Dm(W\Sub{VI})=\Square{3.27\Log{P}-1.79}\delta Z
\label{ZVIPL}
\end{equation}

\smallskip

\noindent
which is only a $0.02$ mag effect for the same assumptions
as above.

By contrast the equivalent formula for $BV$ Wesenheit magnitude
implies a metallicity induced
error of $\delta\Dm\sim-0.21$ under the above assumptions,
similar to the \cite{Stothers88} prediction,
and a much more serious effect.
In drawing conclusions about the effect on HST distance
determination based on $VI$ photometry
we should be aware that these predictions are made
for the standard Johnson $V$- and Cousins $I$-bands.
In fact the central wavelengths of the WFPC2
F555W and F814W filters {\em are} close to these
standard passbands, but the F555W filter, in particular,
is considerably wider than Johnson $V$.
This is likely to result in a greater influence of
line blanketing in that filter, and perhaps therefore,
a better prediction for the metallicity dependence 
would be intermediate between that indicated by 
\eqn\ \ref{ZVIPL} and the equivalent formula for $BI$
data.
Without going into details, we find the mean of these
two to be an even more negligible 
$\delta\Dm\sim-0.01$ mag for the $\delta Z=0.008$ we
have been using for comparison.

How do these predictions 
compare to empirical attempts to quantify the metallicity dependency
of the \PL\ relations?
To date, such investigations have proved
controversial.
Freedman and Madore (1990, hereafter FM90) found no evidence 
for a metallicity
effect in their multicolour 
data for Cepheids in three fields in M31.  
However, \cite{Feast91} and \cite{Stift95} have argued
that this test is not sufficiently sensitive to rule out
an effect of the size predicted by Stothers (1988, hereafter S88).
\cite{Tanvir92}, from an independent analysis of M31 Cepheids, in which
the reddening was estimated from the locus of the main
sequence stars, found that the S88 metallicity correction 
{\em did} give a consistent distance modulus when applied separately
to $B$ and $V$ data.
\cite{Gould94} reanalyzed 
the FM90 data set, taking account of the correlations among
the $BVI$ magnitudes, and claimed to find, contrary to the
original conclusion, a significant
effect of $\delta\Dm=-(0.56\pm0.20)\delta [{\rm Fe/H}]$,
which amounts to $\delta\Dm=0.17$ for the difference 
between the galaxy and the LMC.
Most recently, \cite{Sasselov96}  have analyzed a large
sample of $\sim500$ LMC and SMC Cepheids from the EROS microlensing
experiment.
The large number of variables and high quality of the 
data allows them to simultaneously
determine the differential distance, reddening and 
metallicity effects. They conclude that a fairly large change in distance
modulus of 
$\delta\Dm=-25.4^{+6}_{-12}\delta Z$ would occur for the HST $VI$
data, \ie\ a correction of 0.11--0.25 mag is required 
for $\delta Z=0.008$.
They point out that this correction would go some way
to bringing the individual \H0\ estimates based on the 
distances to
IC4182, M96
and M100 into agreement.
This is much larger than the theoretical prediction given 
above by \eqn\ \ref{ZVIPL}, and,
if correct, implies a problem with 
one of the assumptions
made above or with the CWC93 models
themselves, possibly with the 
M--L relation at low metallicity
(D. Sasselov priv. comm.) or the use by CWC93 of LAOL rather than
more modern opacities.

Clearly this is an area where further work is required to
refine both the observational and theoretical results.
The \kp\ group are attempting an empirical study
of the metallicity dependence in M101 (Kennicutt \etal\ 1995)
which, since they are using HST, at least removes the uncertainties 
associated with  photometric transforms.
For the present, then, the best empirical metallicity
calibration comes from the EROS data, but this seems
to disagree with the existing theory (at least under
the assumptions made here), and itself has a high
internal uncertainty.
The most secure distances therefore will 
be for Cepheid fields with metallicities similar to the LMC,
and we may take $\sigma\Sub{Z}=0.06$ mag as a typical uncertainty
with the proviso that high metallicity galaxies in particular,
will give more uncertain distances.

\section{Future directions}

From the above deliberations it should be
clear that there are several important sources
of systematic uncertainty which need to
be addressed.
Not least of these are the various WFPC2 calibration issues,
indeed there would be  merit in obtaining a WFPC2
photometric calibration specifically for Cepheids
if we want to achieve the highest accuracy.
In addition, if we are to continue to use  the LMC Cepheids
as the primary calibration, then its distance 
needs to be tied down more precisely.  
In this regard, results from the microlensing 
surveys and further ground based observations
of long-period LMC Cepheids will be valuable;
these are already under way at SAAO (D. Laney and
J. Caldwell, priv. comms.) and by the \kp\ team
(Kennicutt \etal\ 1995).  The more local
scale within the Milky-Way is of relevance
for some of the LMC distance determinations
and observations of galactic Cepheids,
the results from the Hipparcos mission
and improvements to the Baade-Wesselink
method (see Krockenberger \etal\ 1996)
will be important here.

In the remainder of this section we highlight
two new developments which will hopefully
lead to better Cepheid distance estimates in the near future.

\subsection{Moving further to the infrared}

There are numerous advantages in observing Cepheids
further in the infrared,
and these have been exploited
from the ground for
nearby galaxies (\eg\ Jacoby \etal\ 1992).  In particular,  dust extinction
is much reduced, there is less sensitivity to metallicity 
variations, and
the spread around the mean \PL\ relation is smaller
than in the optical
(\eg\ McGonegal \etal\ 1982; Laney \& Stobie 1994).
As an illustration of the benefit of reduced
reddening corrections, the true distance
modulus determined by combining $V$- and $H$-band magnitudes
is given by:

\begin{equation}
\Dm=\Mod{AH}-0.24(\Mod{AV}-\Mod{AH})
\label{Hmethod}
\end{equation}
\smallskip

\noindent
which should be compared with \eqn\ \ref{method1}.
As shown in figure \ref{leverarm} this results in
significantly lower sensitivity to zero-point photometry
calibration uncertainties.

The installation of the NICMOS infrared camera
on HST in 1997, will provide the exciting opportunity
to revisit some of the HST-observed galaxies again
to obtain $H$-band magnitudes for the known Cepheids.
Furthermore, the small amplitudes in the  $H$-band
means that only one or two  epochs are required
for  these observations, which makes them very good value
in terms  of telescope time!

\subsection{Maximum light relations}

Cepheid \PL\ relations based on maximum as well as mean
light were presented by
\cite{ST68}.
Recently the idea has been resurrected (Simon, Kanbur and Mihalas, 1993),
motivated by theoretical considerations,
and it has been shown empirically
that the \PL\ relations at maximum light have somewhat lower
dispersion than at mean light
(Kanbur and Hendry, 1996).
Observationally, working  at maximum light removes problems of obtaining
good photometry through the minimum of a Cepheid's cycle, which
is difficult close to the magnitude limit 
especially in crowded fields.
The drawback of the method is that 
it requires careful accounting of the biases 
introduced by sparse sampling of the light curve and
the effects of noise.
It appears that these can be addressed by use of simulations
and therefore \PL$_{\rm max}$ relations hold
considerable promise.

\section{Conclusions} 
\label{conclusions}

We have seen that
the ``standard'' method of extinction correction for
Cepheids is equivalent
to fitting a \PL\ relation to their reddening-free magnitudes.
Because the reddening correction
also accounts for much of the intrinsic colour
dependence of Cepheid magnitudes, this
distance indicator has a very small intrinsic dispersion.
Thus precise results can be obtained for even small
samples of Cepheids, providing the samples are
``clean''.
Bias effects are complicated but likely to be small,
as is contamination from non-Cepheids.
Crowding errors may be a more serious problem
at faint magnitudes,
but all of these can be addressed by simulations and
alleviated if we
impose a conservative lower \log{P} cutoff.

By using the reddening of the Cepheids themselves we
avoid the large uncertainties 
associated with any other way of estimating extinction.
The flip-side of this is that this distance indicator
is quite sensitive to systematic calibration
uncertainties.
With this in mind we have produced revised,
and apparently robust, new
calibrations of the $V$-band,  $I$-band 
and $W\Sub{VI}$ \PL\
relations based on a careful analysis of a
sample of 53 LMC Cepheids.
The formal uncertainty on the zero-point of the $W\Sub{VI}$
relation is only 1\% in distance.
These new relations highlight the calibration
sensitivity since a fairly small change from the
old $I$-band relation means that the new calibration
gives true distance moduli which are approximately 0.1 mag
closer.
The effect of the change will be less for the 
Sandage \etal\ SNIa calibration program (\eg\ Sandage \etal\ 1996),
in those cases
for which the extinction to the 
Cepheids was assumed to be the same as that to the SN.

Finally,
it is instructive to revisit the error budget which was
summarized by \eqn\ \ref{budget}.
The first term, due to the intrinsic errors, will vary from
sample to sample, but because larger samples of Cepheids also
tend to imply more crowded fields and uncertain photometry, it is
reasonable to take a representative figure of 
$\sigma_{\rm internal}/\surd{n}=0.1$ mag.
From section 2, we estimate the next two terms contribute
$0.11^2$, whilst for the fourth term we shall assume a
conservative value of $\sigma\Sub{R}=0.4$
and a typical value of $\delta(\Mod{AV}-\Mod{AI})$ of 0.05 mag.
From section 3, we take $\sigma\Sub{\rm PL}=0.1$ mag, 
which is just the adopted error on the LMC distance since
the internal error on the fit to the Wesenheit magnitudes
is only 0.02 mag and any photometric zero-point
uncertainties on the ground-based 
measurements have been included in $\sigma\Sub{V}$
and $\sigma\Sub{I}$.
In section 4 we saw that the metallicity question is still
unsettled, so we assume a contribution of $\sigma\Sub{Z}=0.06$ mag 
for an average
HST observed galaxy, although of course this uncertainty would
be removed for a galaxy which could be shown to be close to
LMC metallicity.
Finally, $\sigma_{\rm systematic}$ remains poorly determined, but we
shall adopt a value of 0.05 mag, since these uncertainties will tend
to be small and correlated so as not to affect the reddening correction.
In total, then, we find a typical uncertainty of $\sigma_{\rm total}=0.2$ 
mag which amounts to $\sim$10\% in distance.

We have attempted to take reasonable account here of 
all the significant sources of error.
That such precision can be achieved at distances of
20Mpc is a testament to
the exceptional capabilities of the HST.
It is clear, however, that this uncertainty is dominated
by systematic errors, and the value of the whole
sample of HST galaxies can be enhanced considerably 
if these systematics are addressed.
In particular, improved calibration of WFPC2,
specifically those aspects relating to the Cepheid observations;
better understanding of the role of metallicity variations on 
Cepheid properties;
more precise determination of the LMC distance; 
and infrared observations of known Cepheids with NICMOS
are all highly desirable.
With these improvements, and given the many parallel
developments in secondary distance indicators,
the goal of estimating \H0\ to better than 
10\% looks to be achievable.

\begin{acknowledgments}

I would like to thank Dimitar Sasselov, Abi Saha, Dave
Laney, John Caldwell, Wendy Freedman, Barry Madore
and Mike Feast for useful
discussions and communications,
and a security man at Heathrow airport for helping
me feel that at least someone was interested (see
appendix B).
Special thanks go to my collaborators on various
Cepheid related projects:
Tom Shanks,
Martin Hendry, Shashi Kanbur, Shaun Hughes,
Harry Ferguson, David Robinson and
Robin Catchpole for many insightful discussions.

\end{acknowledgments}

\appendix
\section{Transforming $V$ to $I$ light curves}

\begin{figure}[th]
\centerline{\psfig{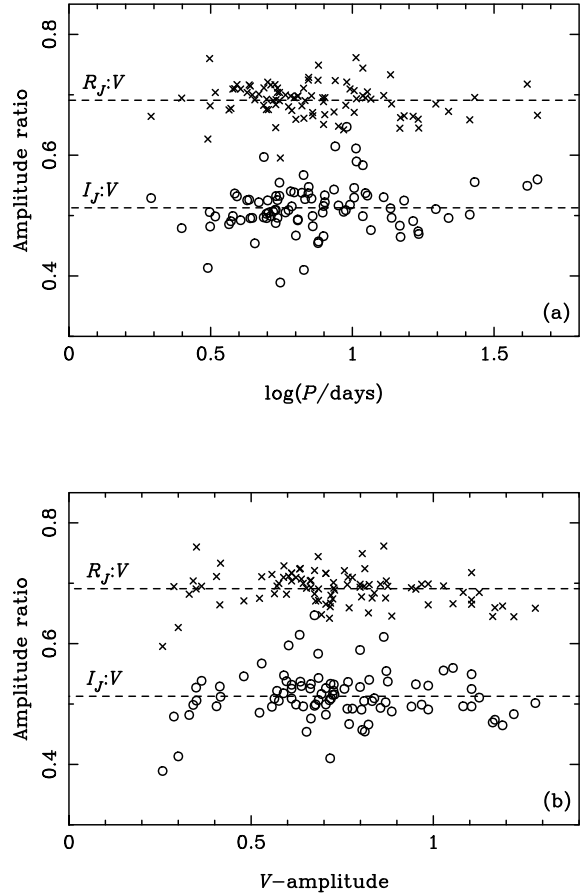}}
\caption{
These two plots show how the ratio of Cepheid 
amplitudes in
$R\Sub{J}$ (crosses) and $I\Sub{J}$ (circles) to the 
amplitude in $V$ depends on (a) $\Log{P}$
and (b) $V$-amplitude (max-to-min) itself.  Apparently there is little
dependence.
The data are for 90 galactic Cepheids taken
from the sample  of \cite{MB85}, from which all type II Cepheids and those
noted as having companions have been removed.
The dashed lines show the means in each case.
}
\label{amprats}
\end{figure}

HST Cepheid studies have all used $V$-band light curve 
shape information to predict the $I$-band light curve 
in order to reduce the number of $I$ epochs which 
are necessary to determine \imm{I}.
The most important factor is the reduction in amplitude between
$V$ and $I$.
We can investigate this empirically with the
large sample of well-studied galactic Cepheids
analyzed by \cite{MB85}.
In figure \ref{amprats} we plot the ratios of the amplitudes in 
Johnson $R\Sub{J}$- and
$I\Sub{J}$-bands to the amplitudes in $V$.
This shows that the ratio is very nearly independent
of $\Log{P}$ and amp$(V)$.
Since F814W, and Cousins $I$,  are intermediate between
Johnson $R\Sub{J}$ and $I\Sub{J}$, we recommend use of a ratio 
${\rm amp}(I):{\rm amp}(V)=0.6$.

It is interesting to note that this amplitude ratio is almost
identical to the reddening ratio, $A\Sub{I}/A\Sub{V}=0.59$.
This suggests that random phase $VI$ observations of Cepheids of
known period should give precise estimates of true distance
modulus after  reddening correction.
However, we note that the 
small phase shift and change of light curve shape in
transforming between the bands will lead to some
additional noise.

\section{A funny thing happened on the way to this meeting}

Traveling to this meeting the author passed through London Heathrow airport
and was stopped by a security guard for a standard
interview, which proceeded something like this:

{\it SG}: ``What is the purpose of your trip to America?''

{\it NRT}: ``I'm going to a conference in Baltimore on the
age of the universe.''

{\it SG}: ``That sounds interesting, do you think you'll decide on an answer
or is it something we don't have much idea about yet?''

{\it NRT}:
``Well, it's very likely we {\em won't} all agree with each other, but
I think we're closer to an answer than you might imagine.''

{\it SG}:
``Oh really, I thought there was a problem with the ages of 
the globular clusters.''!!

\medskip
The moral of this story, I think, is that there are people
out there, even airport security guards, who are really
interested in the big questions about the universe which
we are trying to answer....either 
that or these people receive exceptionally thorough
training on how to catch out unsuspecting academics!
\lower3pt
\hbox
{\small\lower 0.9ex \hbox{${\stackrel{\stackrel{\odot\odot}{o}}{\smile}}$} }

\end{document}